\documentclass[preprint,12pt]{elsarticle}

\usepackage{amsmath,amssymb} \DeclareMathOperator{\sign}{sign}
 \usepackage{graphicx}
\def\al{\alpha}
\def\be{\beta}
\def\ga{\gamma}
\def\de{\delta}
\def\ep{\epsilon}

\def\la{\lambda}

\def\si{\sigma}

\def\mn{{\mu\nu}}

\def\frac#1#2{{\textstyle{{#1}\over {#2}}}}

\def\lsim{\mathrel{\rlap{\lower4pt\hbox{\hskip1pt$\sim$}}
    \raise1pt\hbox{$<$}}}
\def\gsim{\mathrel{\rlap{\lower4pt\hbox{\hskip1pt$\sim$}}
    \raise1pt\hbox{$>$}}}
\def\sqr#1#2{{\vcenter{\vbox{\hrule height.#2pt
         \hbox{\vrule width.#2pt height#1pt \kern#1pt
         \vrule width.#2pt}
         \hrule height.#2pt}}}}

\newcommand{\beq}{\begin{equation}}
\newcommand{\eeq}{\end{equation}}
\newcommand{\bea}{\begin{eqnarray}}
\newcommand{\eea}{\end{eqnarray}}

\usepackage{amssymb}
\journal{Physics Letters B}

\begin{document}

\begin{frontmatter}



\title{Extended Hamiltonian Formalism and Lorentz-Violating Lagrangians}


\author{Don Colladay}

\address{New College of Florida, Sarasota, FL}

\begin{abstract}
A new perspective on the classical mechanical formulation of particle trajectories in 
lorentz-violating theories is presented.
Using the extended hamiltonian formalism,  a 
Legendre Transformation between the associated covariant Lagrangian
and Hamiltonian varieties is constructed.
This approach enables calculation of trajectories using 
hamilton's equations in momentum space and the Euler-Lagrange equations 
in velocity space away from
certain singular points that arise in the theory.
Singular points are naturally de-singularized by requiring the trajectories to be
smooth functions of both velocity and momentum variables.
In addition, it is possible to identify specific sheets of the 
dispersion relations that correspond to specific
solutions for the lagrangian.
Examples corresponding to bipartite Finsler functions are computed in detail.
A direct connection between the lagrangians and the field-theoretic solutions to the 
Dirac equation is also established for a special case.

\end{abstract}

% \begin{keyword}

% \sep Lorentz Violation \sep 



%\end{keyword}

\end{frontmatter}


\section{Introduction}
Computation of classical effective lagrangians that represent motions of wave packets in
theories involving field-theoretic dispersion relations with Lorentz-violating corrections has been of 
considerable interest in the recent literature.
Specifically, the Standard Model Extension (SME) provides a self-consistent framework that
leads to physically viable dispersion relations that incorporate effects due to 
possible Lorentz violation in theories underlying the standard model.
Classical lagrangians arising from these SME dispersion relations have been computed exactly 
in relatively simple algebraic form only for some subsets of the parameters appearing in the general model.
These lagrangians provide a tool for computing classical particle trajectories in a
curved-spacetime background when the background tensors are promoted to space-time
dependent forms that vary slowly over space and time.

The minimal Standard Model Extension (SME) formulated in flat spacetime involves constant background fields that couple to the known particles \cite{kps,ck} through power-counting,
renormalizable, gauge invariant terms.
Extension of the minimal SME theory has developed in several different directions including the
gravity sector\cite{kosgrav} and nonminimal terms \cite{kosmewes}.
Wave packets can be constructed that have specific group velocities which lead to classical particle
trajectories given specific branches of the dispersion relations \cite{brettcol}.  
These paths also follow from Lagrangians computed using a Legendre transformation of the
implicitly defined hamiltonians in the dispersion relations \cite{kr}.
Previous work on SME lagrangians includes computations involving momentum-dependent couplings 
\cite{colmcd1}, non-minimal terms \cite{shreck} and photons \cite{shreck2}.
Much of this work has been related to Finsler geometry \cite{shen1} using either Wick rotations or restrictions
to certain subspaces \cite{kosfins, kosrustso}, or in other contexts 
\cite{erasmo,berger,snow, zheng, vacaru, bonder, bag, gomez, yan, silva, thornberg}, and in analogous
classical systems \cite{ralph}.

The general procedure of computing effective classical Lagrangians leads to a covariant 
Lagrangian when a generalized parametrization is adopted for the four-velocity.
Computation of the associated relativistic hamiltonian yields zero, as is well-known in
standard covariant theories. 
This prevents the inversion of the expression for $p^\mu(u)$ into a formula for $u^\mu(p)$
in a manifestly re-parametrization invariant way and inhibits the natural use of hamilton's equations.

Use of the extended hamiltonian formalism \cite{quantgauge} in which the dispersion 
relation is incorporated into the 
action using a Lagrange multiplier yields a relativistic formulation in which 
hamilton's equations follow naturally.
Singular points occur in both the extended lagrangian and hamiltonian functions when the
associated algebraic varieties fail to be smooth manifolds.
These singular points in the 
lagrangian and hamiltonian varieties are
seen to occur at different points along the particle trajectories.
This means that the full theory involving both varieties can be given a manifold structure and the dual
momentum and velocity variables desingularize each other naturally. 

In this paper, the CPT-violating, spin-dependent $b^\mu$ parameter will be used to illustrate the 
various formulas and definitions as they arise.  
The singular points are identified and the physics of the desingularization is described for
a simple example of a particle trajectory in constant gravitational field.
This case is then generalized to a larger class of bipartite SME dispersion
relations for which the algebraic manipulations are still simple.  

\section{Dirac Equation for $b$-parameter}

The Dirac equation for a fermion in the presence of a CPT- and Lorentz-violating background vector field $b^\mu$
in the minimal SME is \cite{ck}
\beq
\left(i \ga^\mu \partial_\mu - m - b_\mu \ga_5 \ga^\mu \right) \psi = 0.
\label{deq}
\eeq
The corresponding dispersion relation in momentum space can be written as 
$R_T(p) = R_+(p) R_-(p) = 0$, where
\beq
R_\pm(p) = {1 \over 2} \left( p^2-m^2-b^2 \pm 2 \sqrt{(b \cdot p)^2 - b^2 p^2}\right),
\label{hamiltonian}
\eeq
providing an observer covariant (but non-unique...) factorization of the dispersion relation.
There is a map to a related CPT- and Lorentz-violating photon dispersion relation, 
$b^\mu \rightarrow k_{AF}^\mu$, and 
$m^2 \rightarrow m_\ga^2 - b^2$ which is analyzed in detail in \cite{colnoord}.

The plane-wave spinors $u_\pm(p)$ are particle solutions to the (off-shell) Dirac Equation as
\beq
(\not p - m - b_\mu \ga_5 \ga^\mu ) \psi_\pm(x) = 2 R_\pm(p) u_\pm( p) e^{-i p \cdot x}.
\eeq
The above equation reduces upon setting $u_\pm = (\not p + m - \ga_5 \not b) w_\pm$ to
\beq
\ep_{\mu\nu\alpha\be}\si^{\mn} p^\al b^\be w_\pm = \pm 2  \sqrt{(b \cdot p)^2 - b^2 p^2} w_\pm,
\label{sols}
\eeq
which is the condition that $w_\pm$ are eigenstates of the Pauli-Lubanski vector contracted with $b^\mu$.

\section{Classical Lagrangian for $b$-parameter}

The classical lagrangian corresponding to the field-theoretic term in Eq.\ (\ref{deq}) 
is calculated by performing a Legendre Transformation of the associated dispersion 
relation in Eq.\ (\ref{hamiltonian}) and introducing an arbitrary parameterization $\la$, with result \cite{kr}
\beq
{\cal{L_\pm}} = - m \sqrt{u^2} \mp \sqrt{(b \cdot u)^2 - b^2 u^2} ,
\label{lagrangian}
\eeq
where $u^\mu = d x^\mu / d \la$, the invariant product was taken to be flat Minkowskian, and the $b^\mu$ is a constant 
background vector field, with components much smaller in magnitude than $m$ so that the
theory is in a concordant frame \cite{koslehnert}.
This expression may be generalized to curved-spacetime backgrounds by promoting the
constant $b^\mu$ fields to slowly-varying vector fields and the
Minkowski product to a covariant product (ie: $b \cdot u \rightarrow g_\mn(x)b^\mu(x) u^\nu$).
The details of the of the gravity sector SME construction is described in
\cite{kosgrav, kosjay}.

Physically, the two solutions ${\cal L}_+$ and ${\cal L}_-$ have some relation to the helicity solutions 
in Eq.\ (\ref{sols}), although the specific map is not immediately obvious.
The velocity and momentum variables are connected by the definitions 
$u^j/u^0 = - \partial p_0 / \partial p_j$ and
$p^\mu = -\partial {\cal L} / \partial u_\mu$.
When one restricts to regions away
from the singular points, there are two disjoint sheets $R_\pm(p)=0$ which relate the momenta and
velocities on the energy surface in a unique way.  
Note that $u^0(\la)$ is introduced as an arbitrary function adjustable through re-parametrization 
to put the
lagrangian into manifestly covariant form.  This means that there is a gauge type symmetry
that must be fixed to compute the four-velocity.

The Lagrangian is found by solving $R_T(p)=0$ together with setting the total derivative of $R_T$
with respect to $p^j$ equal to zero.  This includes chaining the derivative through the implicit
dependence of $p^0(\vec p)$ due to the constraint $R_T(p) = 0$.
This procedure results in an eighth-order polynomial equation $P({\cal L}) =0$ which can be factored.
The resulting two solutions given in Eq.\ ({\ref{lagrangian}}) are the ones that reduce to the correct classical
form as $b^\mu \rightarrow 0$.
Since the lagrangians in Equ.\ (\ref{lagrangian}) are found through factorization of an eighth-order
polynomial, it is unclear which of the ${\cal L}\pm$ functions are in correspondence with the sheets 
$R_\pm = 0$.
The extended hamiltonian formalism that follows will solve the issue of non-invertibility
of $p(u)$ and provide an explicit connection between the signs chosen in Eqs.\ ({\ref{hamiltonian}})
and ({\ref{lagrangian}}).

\section{New covariant variables for dispersion relation}

A naturally defined observer-covariant four-vector (where the derivatives are taken as if $p^0$ and $p^j$
were independent, or so-called `off-shell' derivatives...) is given by the expression
\beq
w^\mu_\pm = {1 \over m}{\partial R_\pm \over \partial p_\mu} = {p^\mu \over m} 
\pm {(p\cdot b) b^\mu -b^2 p^\mu \over m \sqrt{(b \cdot p)^2 - b^2 p^2}}.
\eeq
A short calculation yields the remarkably simple relation
\beq
R_\pm = m^2 (w_\pm^2 - 1),
\eeq
indicating that the dispersion relation takes the conventional form in terms of these new variables.
This relation was first noticed in \cite{colnoord} where the massive-CPT-violating photon 
dispersion relation takes a similar form.
Inverting for the momentum gives
\beq
p^\mu = m w^\mu_\pm \mp \ep {(w_\pm \cdot b) b^\mu - b^2 w_\pm^\mu \over \sqrt{(b \cdot w_\pm)^2 - b^2 w_\pm^2}},
\eeq
where $\ep = \sign \left( \sqrt{(b \cdot p)^2 - b^2 p^2} \mp b^2 \right)$ 
is a sign factor required near the singular points to obtain the correct relation. 
This looks very similar to the expression for $p^\mu$ in terms of $u^\mu$ computed using the
Lagrangian in Eq.\ ({\ref{lagrangian}})
\beq
p^\mu = - {\partial {\cal L}_\pm \over \partial u_\mu} = {m u^\mu \over \sqrt{u^2}} \pm {(u \cdot b) b^\mu - b^2 u^\mu \over \sqrt{(b \cdot u)^2 - b^2 u^2}}.
\eeq
Note that the $\pm$ signs are flipped due to the reversed notation used for ${\cal L}_\pm$.
In fact, one can see that the two four-velocity parameters $u^\mu$ and $w^\mu$ 
are related by choosing the explicit parametrization
for $u^\mu$ such that $u^2=1$.

This can also be seen by application of the chain rule to the derivative of the dispersion relation 
with respect to $p^j$ used to compute the Lagrangian,
\beq
{dR_T \over dp_j} = {dR_+ \over d p_j} R_- + {dR_- \over d p_j}R_+ = 0.
\eeq
on-shell, on the sheet where $R_- = 0$ (away from the singular point so that $R_+ \ne 0$) we have 
\beq 
{dR_- \over d p^j } = {\partial R_- \over \partial p_0}{\partial p_0 \over \partial p^j} + {\partial R_-
\over \partial p^j} = 0,
\eeq
or
\beq
{w_-^j \over w_-^0} = {u^j \over u^0}, \quad {\rm or} \quad w_-^\mu = {w_-^0 \over u^0} u^\mu = (e_-)^{-1} u^\mu,
\eeq
indicating that $w_-^\mu$ is in fact the velocity four-vector, up to some scalar multiple $e_-(\la)$.
An analogous equation holds for $w^j_+$, involving the introduction of another function $e_+(\la)$.
Defining the Lagrangian as ${\cal L} = - p_\mu u^\mu$ and matching to Eq.\ (\ref{lagrangian}) 
fixes the relation $e_+ = e_- =\sqrt{u^2} $, at least away from the singular points.

\section{Extended Hamiltonian Formalism}

It is useful to `free up' the definition of $e(\la)$ as an auxiliary function to make the four components
of momentum linearly independent using the extended hamiltonian formalism, originally due to Dirac \cite{quantgauge}.  Doing so yields a modified form of the 
candidate \cite{fn1} action functionals as
\beq
S^*_\pm = - \int \left[ m e^{-1} u^2 \pm \sqrt{(b\cdot u)^2 - b^2 u^2} + 
{e \over m} R_\mp (p,x) \right ] d \lambda, 
\label{action}
\eeq
where $ e(\xi \lambda) = \xi e(\lambda)$ is a homogeneous function of degree one to ensure re-parametrization 
invariance of the modified action and $R_\mp(p,x)$ is an appropriate hamiltonian constraint function
that vanishes when the equations of motion are satisfied.  
Note that we take $e_- = e_+ = e$ since this function can be interpreted as a metric on the 
world-line and should be the same for different spin particles if they are allowed to interact.
By writing the action in the form 
\beq
S^* =  \int \left[ - p^\mu u_\mu - {\cal H}^* \right] d \lambda = \int L^* d \lambda,
\eeq
the extended hamiltonian can be 
identified as 
\beq
{\cal H}^*_\pm =  -{e \over m} R_\mp (p,x).
\label{genham}
\eeq
Note that this hamiltonian is zero when the constraint is satisfied (`on-shell'), as is expected for
relativistic systems that are generally covariant (re-parametrization invariance in this case...).
If the constraint is written in terms of $R_\pm(p,x)  = 
{m^2 \over 2} (w_\pm^2 - 1) = {m^2 \over 2 e^2} ( u^2 - e^2)$, then the lagrangian
becomes
\beq
L_\pm^*[u^\mu,x,e] = -{m \over 2 e} u^2 \mp  \sqrt{(b \cdot u)^2 - b^2 u^2} - {e m \over 2}.
\eeq
Variation of this lagrangian with respect to $e$ gives the condition $e = \sqrt{u^2}$, reducing to
the original lagrangian of Eq.\ ({\ref{lagrangian}}) when $e$ is eliminated.
Note that the functional form of the lagrangian is independent of the choice of 
$R_+(p)$ or $R_-(p)$ in Eq.\ (\ref{action}).
The conjugate momenta are now
\beq
p^\mu ={\partial L^*_\pm \over \partial u_\mu} =  {m u^\mu \over e} \pm {(u \cdot b) b^\mu - b^2 u^\mu \over \sqrt{(b \cdot u)^2 - b^2 u^2}},
\label{mom}
\eeq
which is now invertible for $u^\mu(p)$ as
\beq
u^\mu = {e \over m} \left( p^\mu \mp {(b \cdot p)b^\mu - b^2 p^\mu \over \sqrt{(b \cdot p)^2 
- b^2 p^2}}\right),
\label{vel}
\eeq
provided the determinant of the Hessian of $L^*$ with respect to the velocity is nonzero.  
The Hessian is computed as
\beq
h^{\mn}_L = - {\partial^2 L^*_\pm \over \partial u_\mu \partial u_\nu} = {m \over e} \eta^\mn
\mp {b^2 \over ((b \cdot u)^2 - b^2 u^2)^{3/2}} T^\mn(u),
\eeq
with
\beq
T^\mn (u) = ((b \cdot u)^2 - b^2 u^2) \eta^\mn + b^2 u^\mu u^\nu + u^2 b^\mu b^\nu 
- (b \cdot u)(b^\mu u^\nu + b^\nu u^\mu).
\eeq
The determinant is then computed by acting on a linearly independent basis of eigenvectors as
\beq
det(\eta \cdot h_L) = \left( {m \over e} \left[ {m \over e } \mp {b^2 \over \sqrt{(b \cdot u)^2 
- b^2 u^2}}\right] \right)^2,
\label{hessl}
\eeq
valid when $b$ and $u$ are not parallel.
When $b$ and $u$ are parallel, the Lagrangian becomes independent of the Lorentz violation
parameter producing singular behavior.
An additional source of singular behavior is due to the vanishing of the determinant for the 
upper sign when $e$ happens to satisfy
\beq
e = {m \sqrt{(u \cdot b)^2 - b^2 u^2} \over b^2 },
\eeq
which can happen for some physical values of $u$ when $b$ is time-like.
Evaluation of the momentum at these points yields
\beq
p^\mu \rightarrow {(u \cdot b)b^\mu \over \sqrt{(u \cdot b)^2 - b^2 u^2}},
\eeq
which is degenerate for some set of nonzero velocity four-vectors.
For example, imposing the equations of motion fixes $e^2 = u^2$ and imposing standard
parametrization so that $u^2 = 1$ implies the determinant vanishes for three-velocities 
satisfying
\beq
\gamma^2 (b^0 - \vec b \cdot \vec v)^2 = b^2(1 + {b^2 \over m^2}),
\eeq
where $\gamma = 1/\sqrt{1 - \vec v^2}$ denotes the standard relativistic factor.
Solutions to this equation occur for values of the velocity of order $b$, for example, if
$b^\mu = (b_0,0,0,0)$, then the Lagrangian is singular
on the sphere determined by $| \vec v| = b_0 / \sqrt{m^2 + b_0^2}$, and
$p^\mu = (\sqrt{m^2 + (b_0)^2},0,0,0)$.
Note that the direction of $\vec v$ can be used to characterize the trajectory
at points where the momenta are zero.

The corresponding extended hamiltonian functions are 
\beq
{\cal H}^*_\pm = -{e \over 2m} \left(p^2 - m^2-b^2 \mp 2 \sqrt{(b \cdot p)^2 
- b^2 p^2}\right),
\eeq
with Hessian matrix 
\beq
h^{\mn}_H =  -{\partial^2 {\cal H}^* \over \partial p_\mu \partial p_\nu} = {e \over m}\left[ \eta^\mn
\pm {b^2 \over ((b \cdot p)^2 - b^2 p^2)^{3/2}} T^\mn(p) \right],
\label{hessh}
\eeq
with determinant
\beq
det(\eta \cdot h_H) = \left({e \over m}\right)^4 \left[ 1 \pm {b^2 \over \sqrt{(b \cdot p)^2 
- b^2 p^2}}\right]^2,
\eeq
which vanishes on a set in momentum space, complimentary to the one in velocity space.
This is useful since Hamilton's equations relate derivatives of the extended hamiltonian to 
the velocity covariantly as
\beq
{\partial {\cal H}^*_\pm \over \partial p_\mu} = - u^\mu, 
\quad {\partial {\cal H}^*_\pm \over \partial x^\mu} =  \dot p_\mu.
\label{hameq1}
\eeq
Note that the second equation becomes useful when the background metric varies from 
a flat Minkowskian one.
Note that it is crucial that the four components $p^\mu$ be varied independently in the proof
that hamilton's equations hold, which is now possible due to the inclusion of the auxiliary 
$e$ parameter into the theory.

When the extended hamiltonian is expressed in terms of the velocity, it takes the
conventional form
\beq
{\cal H}^*_\pm = - {m \over 2 e}(u^2 - e^2).
\label{hamu}
\eeq
A corresponding expression for the lagrangian in terms of the momentum variables is
\beq
L^*_\pm = -{e \over 2 m}(p^2 + m^2 + b^2) .
\eeq
It is curious that both the extended hamiltonian and lagrangian take the conventional 
relativistic form when expressed in terms of the "wrong" variables.
Note that the above formulas serve to define the theory as a one-to-one Legendre transformation
provided a certain singular set at low velocities and momenta are avoided.  
Mathematically, this region corresponds to points where $D(p) \equiv \sqrt{(b \cdot p)^2 - b^2 p^2}$
and $D(u) \equiv \sqrt{(b \cdot u)^2 - b^2 u^2}$ fail to be in one-to-one correspondence.
These functions are related by
\beq
\ep_H D(p) - \ep_L {m \over e} D(u) = - b^2,
\label{drel}
\eeq
where $\ep_L$  ($\ep_H$) is the sign chosen in $L^*_{(\ep_L = \pm)}$  ($H^*_{(\ep_H = \pm)}$).
The hessians are badly behaved when $D(p) \sim b^2$, indicating a lack of convexity in a 
small region near the points where the determinants vanish.
Outside of this singular region, one is free to choose one of the $\pm$ signs in $L^*_\pm$ and 
use the corresponding extended hamiltonian ${\cal H}^*_\pm$, and the relation in Eq.\ (\ref{drel})
becomes 
\beq
D(p) - {m \over e} D(u) = \mp b^2,
\eeq
and the equations relating the momentum and velocity given in Eqs.\ (\ref{mom}) and (\ref{vel})
are one-to-one providing a well-defined Legendre transformation on an open convex subvariety of
the solution space.
This can be seen directly through the formula
$(h_L)^{\mu \al} \cdot (h_H)_{\al \nu} = \de^\mu_{~\nu}$, which follows directly from the chain rule.
Within the singular region, it is not possible to use a single global sign choice to define the
action and
some procedure is required to handle the signs of the functions appearing in
Eq.\ (\ref{action}) more carefully.  
This topic is addressed next.

\section{Behavior Near Singular Points}

The determinants of the hessian matrices in Eqs.\ (\ref{hessl}) and (\ref{hessh}) vanish
when either
$D(u) = \ep_L {e b^2 \over m}$ or $D(p) = \ep_H {b^2}$.  In addition, when $det(\eta \cdot h_H) = 0$, the corresponding velocity function $D(u)$ vanishes and $det{(\eta \cdot h_L)}$ diverges to either $\pm \infty$.  An important observation arising from Eq.\ (\ref{drel}) is that
it is not possible for both $D(u)$ and $D(p)$ to simultaneously vanish (provided $b^2 \ne 0$).

In order to handle the relative sign choices in Eq.\ (\ref{action}), the expressions for 
the extended lagrangian $L^*$ and hamiltonian ${\cal H}^*$ can be
re-expressed in terms of the zero set of the following polynomials (which define algebraic varieties)
\beq
f_L[L^*,u^\mu,e] = \left( L^* + {m \over 2 e} (u^2 + e^2) \right)^2 - D^2(u) = 0,
\eeq
and
\beq
f_H[{\cal H}^*,p^\mu,e] = \left( {\cal H}^* + {e \over 2 m}(p^2 - m^2 - b^2)\right)^2 
- ({e \over m})^2 D^2(p) = 0
\eeq
The gradients of these functions are nonzero provided $D(u) \ne 0$ and $D(p) \ne 0$, indicating that
the corresponding varieties are smooth everywhere except at singular points where either 
$D$-function vanishes.
Derivatives of $L^*$ and ${\cal H}^*$ and the corresponding Legendre transformation 
can therefore be defined implicitly on these varieties  everywhere except at the singular points.  

At the singular points, the lagrangian variety can be formally blown up using an 
auxiliary set of variables as was 
demonstrated in \cite{desing} using the non-extended Lagrangian formalism.
Here, it is demonstrated that the momentum-space variables can be used to parametrize
the variety in velocity space near the singular point $D(u) = 0$, naturally desingularizing it.  
An $n-2$ dimensional sphere of
momentum values degenerate to the same velocity value at the singular point, so by retaining 
this information, it is possible to define smooth paths on the Lagrangian variety through the singular
points by observing that the momentum variables are continuous due to Hamilton's equations.  
A symmetric procedure can be used to handle paths going through the singular points $D(p) = 0$
on the hamiltonian variety.

As an example, consider a particle moving vertically in a region of constant gravitational field 
near the surface of the Earth for which
the metric is given by
\beq
d \tau^2 \approx \left(1 + {2 g} z \right) dt^2 - \left ( 1 - {2 g} z \right) dz^2,
\eeq
using clock time $t$ at the surface and the height from the surface $z << R$ as coordinates,
with  $b^\mu = (b^0,0,0,0)$, and constant $b^0$ in this coordinate system.
Using the proper time as a parameter gives the Euler-Lagrange equation that follows from
Eq.\ ({\ref{lagrangian}}) as
\beq
m {d v \over d t} \mp b^0 {d \over d t} \left( {\partial |v| \over \partial v} \right) = - mg,
\eeq
where the motion is taken to be non-relativistic and $v = dz/dt$ for motion along the vertical direction.
Note that the singular point at $v=0$ is evident as the correction term is not defined there, but vanishes everywhere else.
The geodesic through the singular point can be determined uniquely by examination of Hamilton's
equation of motion in Eq.\ (\ref{hameq1}) which reduces to
\beq
\dot p_z \approx -mg \left( 1 \pm  2 {b^0 |p^z| \over m^2} \right),
\eeq
proving that the momentum variables remain continuous through the trajectory near the singular point.
Examination of 
\beq
p^z \approx mv \mp b^0 {\partial |v| \over \partial v},
\eeq
demonstrates that the particle must transition from $L_\pm$ to $L_\mp$ as it passes through 
the singular point at $v=0$ as the the term $\partial |v| / \partial v$ flips from $+1$ to $-1$ during the transition.
Physically, this trajectory can be described as a particle with spin-up rising to its apex and falling
again with spin remaining up.  During this process, the velocity helicity changes sign in passing
through the apex of the trajectory while the energy and momentum remains continuous.
If instead, the particle remains on the same lagrangian sheet, the spin would have to flip at 
the top requiring a discontinuous change in the energy and momentum in violation of
hamilton's equations.

The corresponding singular point in momentum space occurs when $p^z = 0$, which occurs 
at $v = b^0 / m$ when using $L_+$.
In this case, the Euler-Lagrange equation requires the velocity to be continuous through the singular point with the implication that
the particle transitions from ${\cal H}_- = 0$ to ${\cal H}_+ = 0$ as it passes through this velocity.
In this case, it is the momentum helicity that flips sign.
This means that the action given in Eq.\ (\ref{action}) needs to be modified in the neighborhood
of the singular point so that the appropriate hamiltonian is paired with the chosen lagrangian
and visa-versa.  

\section{Generalization to bipartite case}

The above case for $b^\mu$ parameter can be put into a compact form using 
$s_\mn = b_\mu b_\nu - b^2 g_\mn$,
so that $D_s(p) = \sqrt{s_\mn p^\mu p^\nu}$.
It turns out that a class of easily solved Legendre transformations exists when the matrix $s^\mn$
is arbitrary but still satisfies the special condition $s^2 = -\zeta s$ \cite{kosrustso}.
The corresponding momentum space constraint is
\beq
{1 \over 4} (p^2 - m^2 - \zeta)^2 -  s_\mn p^\mu p^\nu = 0,
\eeq
which can be factored into the form $R_+ R_- = 0$, analogous to the $b$-case.
This gives rise to the extended Hamiltonians
\beq
{\cal H}^*_\pm = -{e \over 2 m} \left( p^2 - m^2 - \zeta \mp \sqrt{p \cdot s \cdot p} \right).
\eeq
Computation of the four-velocity gives
\beq
u^\mu = - {\partial {\cal H}^* \over \partial p_\mu} = {e \over m} \left( p^\mu 
\mp {(s \cdot p)^\mu \over \sqrt{p \cdot s \cdot p}}
\right),
\eeq
which gives the same expression as Eq.\ ({\ref{hamu}}).
This simple formula gives some additional insight into why the bipartite form is so special, it
leads to a conventional description of the system when written in terms of alternative variables.
The extended Lagrangian becomes
\beq
L_\pm^*[u,x,e] = - {m \over 2 e} u^2 \mp \sqrt{u \cdot s \cdot u} - {m e \over 2} .
\eeq
singular subspaces occur when either the momenta or the velocity vectors are killed by $s$.
The relation in Eq.\ ({\ref{drel}}) generalizes to 
\beq
\ep_H D_s(p) - \ep_L {m \over e} D_s(u) = - \zeta,
\eeq
where again, the $\ep_H$ and $\ep_L$ are the sign choices used in the extended hamiltonian
and lagrangian functions.
This relation implies that at the singular points where one of the $D$-functions vanishes,
the other one is nonzero in a neighborhood of that point, provided $\zeta \ne 0$.
This means that the momenta variables can be used to de-singularize the velocity variables
and visa-versa as in the $b$-case.

\section{Conclusion}

Using the extended hamiltonian formalism, the classical mechanics implied by
Lorentz-violating dispersion relations in the SME can be implemented using 
both the Euler-Lagrange and hamilton's equations simultaneously.
The formulation is manifestly covariant in that an einbien $e$ is introduced
to free up the variations of the extended hamiltonian with respect to all four momentum components.

In this formailsm, the theory provides an explicit connection between the 
choice of lagrangian and the 
original energy surfaces from which it was derived that can allow for a 
physical interpretation of the states in terms of
the original field theoretic model.  For example, in the $b$ case, it is the eigenstates of the 
Pauli-Lubanski operator contracted with $b$ that determine the energy surfaces and the
associated lagrangian functions.
In addition, the symmetric treatment of variables in velocity and momentum space allows for 
natural de-singularization when the momentum-space variables are used to parametrize the 
velocity space singular points, and visa versa.
As an added benefit, the particle trajectory equations can be formulated directly in momentum
space, thereby removing the necessity to 
first convert to velocity space lagrangians
which can be intractable algebraically in many situations.
This may be particularly useful when considering interacting theories as it is the 
total momentum that is conserved rather than the total velocity.
Successful application of this formalism to non-bipartite SME dispersion relations 
remains an interesting open issue.




\begin{thebibliography}{00}
\bibitem{fn1}
Note that the two sets of $\pm$ signs are correlated as shown over the majority of the
space, however, these signs must be carefully chosen in the vicinity
of singular points as will be discussed later in the paper.
\end{thebibliography}


\end{document}